# Spin-Dependent Limits from the DRIFT-IId Directional Dark Matter Detector

E. Daw [a], J.R. Fox [b], J.-L. Gauvreau [b], C. Ghag [c], L.J. Harmon [b], M. Gold [d], E.R. Lee [d], D. Loomba [d], E.H. Miller [d], A.StJ. Murphy [c], S.M. Paling [a], J.M. Landers [b], M. Pipe [a], K. Pushkin [b], M. Robinson [a], D.P. Snowden-Ifft [b *], N.J.C. Spooner [a], D. Walker [a]

*Corresponding author.

*E-mail address*: ifft@oxy.edu (D.P. Snowden-Ifft)

[a] *Department of Physics and Astronomy, University of Sheffield,  Sheffield, S3 7RH, UK.*

[b] *Department of Physics, Occidental College, Los Angeles, CA  90041, USA.*

[c] *SUPA, School of Physics and Astronomy, University of Edinburgh, Edinburgh, EH9 3JZ, UK.*

[d] *Department of Physics and Astronomy, University of New Mexico, NM 87131, USA.*

**Abstract**

Data are presented from the DRIFT-IId detector operated in the Boulby Underground Science Facility in England.  A 0.8 m$^3$ fiducial volume, containing partial pressures of 30 Torr $CS_2$ and 10 Torr $CF_4$, was exposed for a duration of 47.4 live-time days with sufficient passive shielding to provide a neutron free environment within the detector.  The nuclear recoil events seen are consistent with a remaining low-level background from the decay of radon daughters attached to the central cathode of the detector.  However, charge from such events must drift across the entire width of the detector, and thus display large diffusion upon reaching the readout planes of the device. Exploiting this feature, it is shown to be possible to reject energy depositions from these radon progeny recoil events while still retaining sensitivity to fiducial-volume nuclear recoil events. The response of the detector is then interpreted, using the F nuclei content of the gas, in terms of sensitivity to proton spin-dependent WIMP-nucleon interactions, displaying a minimum in sensitivity cross section at 1.8 pb for a WIMP mass of 100 GeV/c$^2$. This sensitivity was achieved without compromising the direction sensitivity of DRIFT.

## 1. Introduction

Observational evidence from many sources has led to the general acceptance that cold, non-baryonic dark matter forms a large fraction of the energy density of the Universe. A well-motivated explanation of the origin of this material is that it is composed of Weakly Interacting Massive Particles (WIMPs) formed in the Big Bang, a view supported by the observation that the required matter energy density closely matches that calculated from expansion-driven freeze-out of generic weak-interaction annihilations. Dark matter candidates naturally arise in many extensions of the standard model of particle physics, for example in supersymmetric models where the lightest supersymmetric partner is a neutralino [1].

Evidence in support of the dark matter hypothesis may come from indirect measurements, such as measurement of neutrinos from dark matter annihilations in the Sun, or from accelerator searches, such as through searches for supersymmetric particles at the LHC. However, a truly robust signature is given by direct detection of WIMPS interacting with ordinary matter. Due to the smallness of the WIMP cross sections and the difficulty of reducing and predicting backgrounds, measurement of the recoil direction of elastically scattered nuclei is widely regarded as being the most robust direct detection signal [2]. At present low pressure TPCs, such as DRIFT, offer the best technology for providing such measurements [1, 3].

The DRIFT collaboration, through construction of the DRIFT I and II series of experiments (described in detail in [4]) has pioneered the use of low-pressure negative ion gas time projection chambers for this purpose. In [5] and [6] it has been demonstrated that the DRIFT-II detectors are capable of extracting directional information from low-energy WIMP recoils.

Here, an analysis and interpretation of 47.4 days of live time data accrued with DRIFT-IId (the latest DRIFT-II detector) filled with partial pressures of 30 Torr $CS_2$ and 10 Torr $CF_4$ is presented. The use of $CF_4$ is interesting [7, 8] because $^{19}F$ (natural abundance 100%) has a ground state spin-parity of $\frac{1}{2}^+$ originating from an un-paired proton, resulting in a target capable of providing sensitivity to proton spin-dependent WIMP-nucleon interactions. Even with relatively small target mass, with such sensitivity DRIFT-IId is capable of exploring interesting regions of spin-dependent interaction phase space. Furthermore, in the present work, an analysis that is able to reject Radon Progeny Recoil (RPR) events, DRIFT's only known background [9], is presented. Using appropriate neutron calibration exposures, and supported with Monte Carlo simulations, a signal region within which DRIFT-IId was background free but retained sensitivity to nuclear recoils was defined. The use of this background free signal region, and carefully benchmarked WIMP recoil simulations, allowed the calculation of limits on spin-dependent WIMP cross sections.

## 2. The DRIFT-IId detector

The hardware used in this experiment, DRIFT-IId, was identical to DRIFT-IIb [4] with the exception of a redesigned gas input system that allowed arbitrary mixtures of gases to be fed, continuously, into the TPC. A $1.5^3$ $m^3$ low background stainless steel vacuum vessel provided containment for this gas mixture. Within the vacuum vessel were two back-to-back TPCs with a shared, vertical, central cathode constructed of 20 μm stainless steel wires with 2 mm pitch. Two field cages, located on either side of the central cathode, defined two drift regions of 50 cm depth in which recoil events could be observed.  Charge readout of tracks was provided by two MWPCs each comprised of an anode plane of 20 μm stainless steel wires with 2 mm pitch sandwiched between (1 cm gap) two perpendicular grid planes of 100 μm stainless steel wires also with 2 mm pitch.  The potential difference between the grids and the grounded anode planes was -2757 V.  The -30.175 kV central cathode voltage produced a drift field of 549 V/cm.  The mobility of negative $CS_2$ ions in this gas mixture was measured [10] to be 0.57±0.01 $cm^2$atm/Vs which produced a drift speed of 5944 cm/s.  For each MWPC, 448 grid wires ($y$-direction) were grouped down to 8 "lines" which were then pre-amplified, shaped (4 μS shaping time) and digitized. The anode wire signals ($x$-direction) were treated identically except that the electronic gain was half that of the grids.  Eight adjacent readout lines (either anode or grid) therefore sampled a distance of 16 mm in $x$ and $y$.  Voltages on the grid and anode lines were sampled at 1 MHz providing information about the event in the third ($z$-direction) dimension.  The 52 wires at the edges of the grid and anode planes were grouped together to provide veto signals for each MWPC.  Triggering of the data acquisition system (DAQ) occurred when the sum of the anode lines for this experiment exceeded a level of 50 ADC units, equivalent to 24.4 mV.  All lines were digitized with 12 bit resolution from -3000 μS to +7000 μS relative to the trigger. The region bounded by the vetoes and the inner grid planes formed a fiducial volume of 0.80 $m^3$ equating to, with a 30 Torr $CS_2$ and 10 Torr $CF_4$ gas mixture, a target mass of 139 g.  Each side of the detector was instrumented with an automated, retractable, ~100 μCi $^{55}$Fe calibration sources which allowed monitoring of detector gain and functionality.  Lastly, the entire vacuum vessel was surrounded by polypropylene pellets with a thickness of 44 g cm$^{-2}$ as a shield from ambient neutrons [9].

## 3. Data and analysis

Data were collected nearly continuously over 55 days starting in December 2009 during which 47.4 live-time days of data were recorded. These data were subject to an analysis designed to maximize acceptance of nuclear recoil events while rejecting background events. A brief description of the analysis procedure is provided here. All 36 lines of data showed evidence of 55 kHz and 50 Hz noise pickup and harmonics thereof. These were removed with a combination of Fourier analysis and fitting routines. Following the selection of software thresholds (set independently for anode, grid and veto lines) and a region of interest (ROI), each line was analyzed to produce a number statistics. For this analysis the thresholds were set at ~10 standard deviations above the noise with a ROI from -200 μS to 500 μS relative to the trigger. Among the statistics generated were the integral of the voltage with time ($\Sigma$) between baseline crossings if, and only if, the waveform crossed the threshold. $\Sigma$ has been found to be proportional to the integrated charge falling on the MWPC, reported as the number of ion pairs (*NIPs*). The proportionality constant was obtained from $^{55}$Fe (5.9 keV X-ray or 234 *NIPs* per interaction using $W = 25.2$ eV [11]) calibrations done every 6 hours. As described in [9], a careful analysis of this calibration data allowed an estimate of the proportionality constant to better than 1% on an event-by-event basis.

Analysis of the data, employing the above and other event-by-event statistics, was able to identify and remove events inconsistent with nuclear recoils. Briefly, the cuts were intended to remove, in order of decreasing likelihood, large events in which the amplifiers saturated (mostly due to sparks within the MWPC or to ground), alpha particles, events that occurred over the veto regions, small sparks or events happening inside the MWPC, any possible residual gamma-ray induced events, events with pre, post or other side ionization and ringers [9, 12]. Overall the accepted event rate of $130 \pm 2$ events per day. As discussed in [9], these events are consistent with RPRs originating from the 20 μm wire central cathode.

Since this work is focused on spin-dependent WIMP-proton interactions and F is the only nucleus in the gas with significant non-zero spin content, the ionization (*NIPs*) produced by each event was converted, using [13], to F recoil equivalent energy. The distribution of F recoil equivalent energies is shown in Fig. 1. A peak is expected from RPR events, however, as described in [9], the low energy side of the peak was heavily influenced by the cuts described above. DRIFT-IId had an effective threshold of ~20 keV F recoil equivalent energy for this analysis.

## 4. RPR rejection

With the knowledge that the observed events were primarily RPRs, a methodology to discriminate RPRs from WIMP recoils was developed. RPRs originate at the central cathode and therefore experience maximum diffusion in the detector [14]. Thus, due to diffusion, the spatial extent an RPR event might be expected to be, on average, larger than the width of a WIMP recoil event. The most accurate measure of the width of an event was found to be the induced waveform subtracted (IWS) anode sum line rms time of the pulse (*RMST*). A full description of the IWS analysis may be found in [5]. Fig. 2 shows a plot of F recoil equivalent energy vs. the *IWS RMST* for both science run data (in black) and neutron recoil data produced (in red) by a $^{252}$Cf source, discussed below. As discussed in [5], neutron recoils from a $^{252}$Cf source are a good approximation of nuclear recoils from massive WIMPs. As expected, the neutron recoil events had, on average, smaller *IWS RMST*s than the RPR events providing a way of discriminating these two populations. Fig. 3 shows only the science run data along with an acceptance window at small *IWS RMST* in which F recoils are expected but in which none were observed. Limits could therefore be obtained from this data set.

## 5. Recoil calibration and simulation

As discussed in [5], elastic recoils from $^{252}$Cf neutrons provide an ideal calibration data set. Data taken during an exposure of the DRIFT-IId detector to $^{252}$Cf neutrons on February 16[th] 2010, towards the end of the science run, was used for this purpose. A hollow plastic tube of diameter 10 cm was inserted vertically through the shielding centered on the plane containing the central cathode and 10+/-1 cm in from the front door and the pellets inside were removed. The $^{252}$Cf source was then placed on the top of the vacuum vessel inside this hole. The activity of the source at the time of the exposures was 3700±200 neutrons/sec (manufacturer's information and an independent measurement by the DRIFT collaboration [9]). To inhibit gamma ray interactions the source was contained within a cylindrical lead canister of wall thickness 1.3 cm and outer dimensions 5 cm diameter by 11 cm length. 0.819 live time days of data were recorded. After analysis cuts the average rate of accepted events was 4300±100 events per day, 33 times the science run rate.

*GEANT4*, which modeled neutron rates in the DRIFT-IIa detector [9] with several percent accuracy, was modified to simulate the experimental situation described above. Elastic and inelastic interactions with nuclei within the fiducial volume had their recoil types, initial energies, initial direction vectors and initial interaction locations fed into a detector simulation. In this simulation initial recoil energies were converted to *NIPs* using quenching factors calculated by Hitachi [13]. *SRIM-2008.03* [15] was then used to convert the initial recoil energies into nominal ranges. The *SRIM-2008.03* output for lateral and longitudinal straggling were then added onto these nominal ranges. Straight line segments between the initial interaction locations and the randomly determined final locations approximated the motion of the ions in the gas. The negative ions were then uniformly distributed over these line segments, diffused, as per [14], and avalanched onto anode wires with a Polya distribution [16]. To approximate the effect of the amplifiers, each avalanche was convolved with a Gaussian whose height was determined by the gain chain and whose width was determined by the shaping amplifiers. In this way each interaction was converted into a voltage trace on 8 lines of a simulated DRIFT-IId detector. Induced pulses were added to adjacent wires. Signals from the grid were also simulated via induced pulses whose Gaussian width is determined by the geometry of the detector [16]. Finally an approximation to background/amplifier noise was also added to each line. These simulated data were then written out to a file in exactly the same format as data from the real DRIFT-IId detector.

The advantage of this technique is that simulated data could be treated in exactly the same manner as data from the real DRIFT-IId detector. Any biases imposed on data from the real DRIFT-IId detector by the analysis were also imposed on the simulated data. The outputs of the analysis were then compared and adjustments made to the simulation parameters.

Fig. 4 shows the real and predicted F recoil equivalent energy vs. *IWS RMST* plots. For the purposes of limit setting the fraction of events falling within the signal region is the most important prediction. The simulation predicted 226 events to fall in the signal region in good agreement with the 228 +/- 15 observed events.

## 6. WIMP recoil simulation and limits

Having established the accuracy of the simulation what remained was to generate recoils from WIMPs. WIMP velocities were generated with a distribution governed by equations and recommended parameters in [17], i.e. $\rho_D = 0.3$ GeV/c$^2$/cm$^3$, $v_0 = 230$ km/s, $v_E = 244$ km/s and $v_{esc} = 600$ km/s. An assumed WIMP mass then allowed for a distribution of F recoils to be generated. This information was then fed into the calibrated detector simulation generating simulated, WIMP data which was then processed by the analysis code identically to real data. A plot of F recoil energy vs. *IWS RMST* generated from 100 GeV WIMPs is shown in Fig. 5. In this run 9,000 WIMP-F recoil events were generated, 700 events passed all analysis cuts and 118 events fell within the acceptance window. Similar results were then scaled to obtain a 90% C.L. WIMP-nucleus interaction cross section. The procedure outlined in [18] was used to convert the WIMP-nucleus interaction cross-section into a WIMP-proton interaction cross-section for comparison with other experiments.

## 7. Results and discussion

The limits obtained from this procedure are displayed in Fig. 6. Several comments are appropriate. First, none of the other groups' limits use a consistent set of WIMP velocity parameters making comparisons difficult. The parameters for the DRIFT curve are the same as for the PICASSO experiment. Second, this was not a "blind" analysis. For future DRIFT results the procedure for a fully "blind" analysis of the data is now established and will be used. The limits shown in Fig. 6 serve to demonstrate that the limit setting power of the DRIFT-IId detector, despite its low mass, is comparable with the world's best spin-dependent WIMP-proton limits.

## 8. Conclusion

A direct search for weakly interacting massive particles was conducted with the DRIFT-IId detector operating with a gas mixture that provided sensitivity to spin-dependent interactions, and in a mode that retained its ability to reconstruct the direction of nuclear recoils at low energy. A 47.4 live days exposure of $0.8 \text{ m}^3$ of 30 Torr $CS_2$ and 10 Torr $CF_4$ revealed a population of events consistent with recoil decay progeny of radon nuclei located on the central cathode. A technique based on spatial diffusion was used to fiducialize and reject these events. A non-blind analysis of the remaining fiducial volume then allowed the exclusion of proton spin-dependent interaction cross sections displaying a minimum in sensitivity (90% C.L.) at 1.8 pb for a WIMP mass of $100 \text{ GeV/c}^2$. These results demonstrate that future directionally sensitive DRIFT devices will be competitive in the search for dark matter.

## 9. Acknowledgments

We acknowledge the support of the US National Science Foundation (NSF). This material is based upon work supported by the NSF under Grant Nos. 0855933 and 0856026. Any opinions, findings, and conclusions or recommendations expressed in this material are those of the author(s) and do not necessarily reflect the views of the NSF.  We are grateful to Cleveland Potash Ltd and the Science and Technology Facilities Council (STFC) for operations support and use of the Boulby Underground Science Facility.

**Figures**

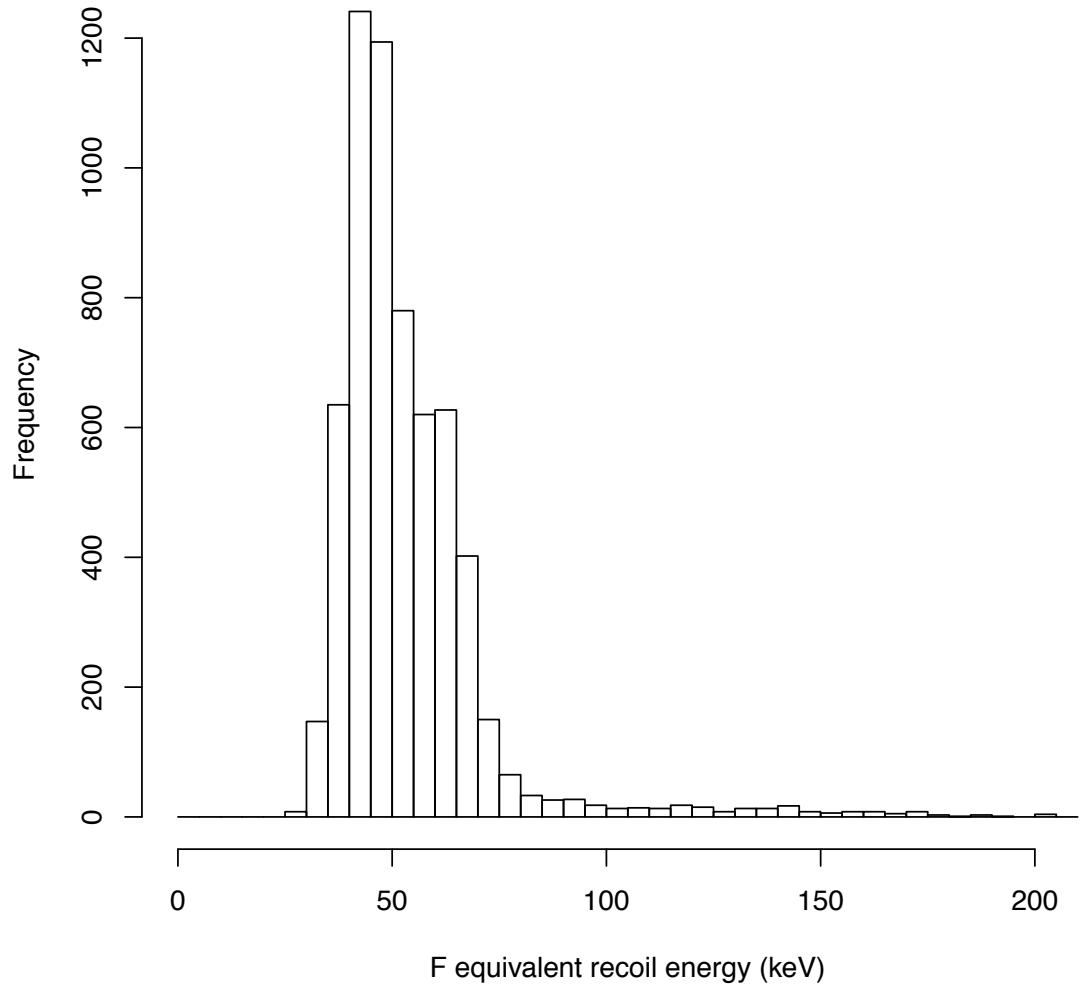

Fig. 1 – The distribution of recoil energies observed during the science run where the measured ionization has been converted into F equivalent recoil energies for convenience.

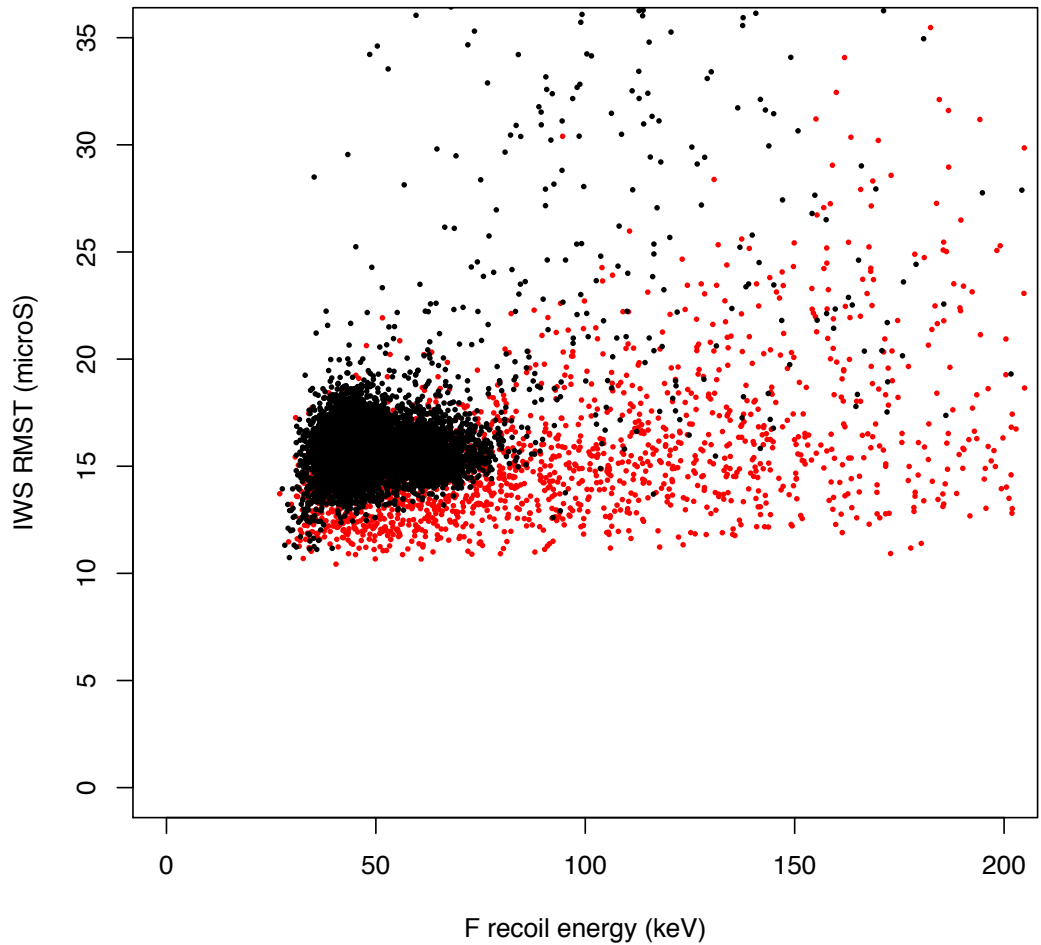

Fig. 2 – A plot of the F recoil equivalent energy vs. the *IWS RMST*, which is a measure of the width of an observed event, for both science run events, shown in black, and neutron recoil events, shown in red. As expected, the neutron recoils events had, on average, smaller *IWS RMST*s than the RPR events providing a way of discriminating these two populations.

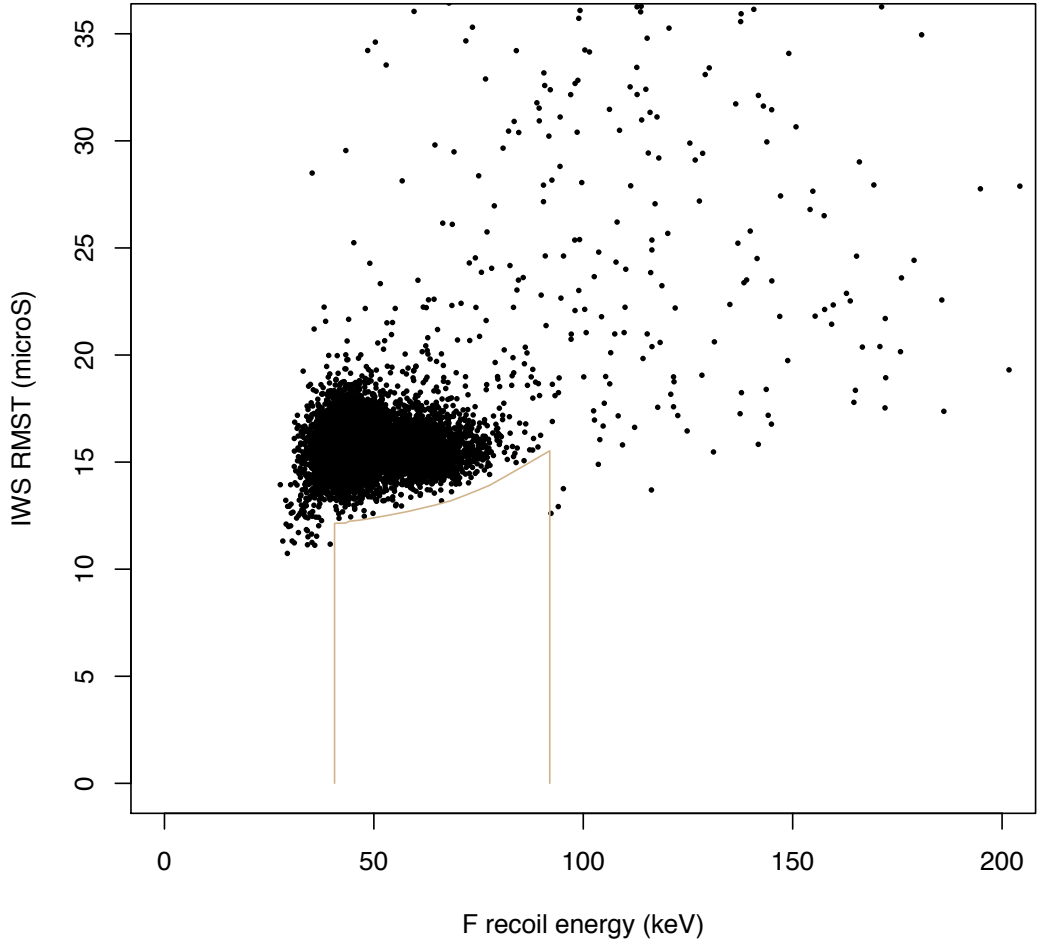

Fig. 3 – A plot of the F recoil equivalent energy vs. the *IWS RMST* for all science run events. An acceptance window obtained after the data was analyzed (i.e. not a blind analysis) is shown in tan. There are no events in this window but, from Fig. 2, WIMP F recoils would be expected there if present in sufficient numbers.

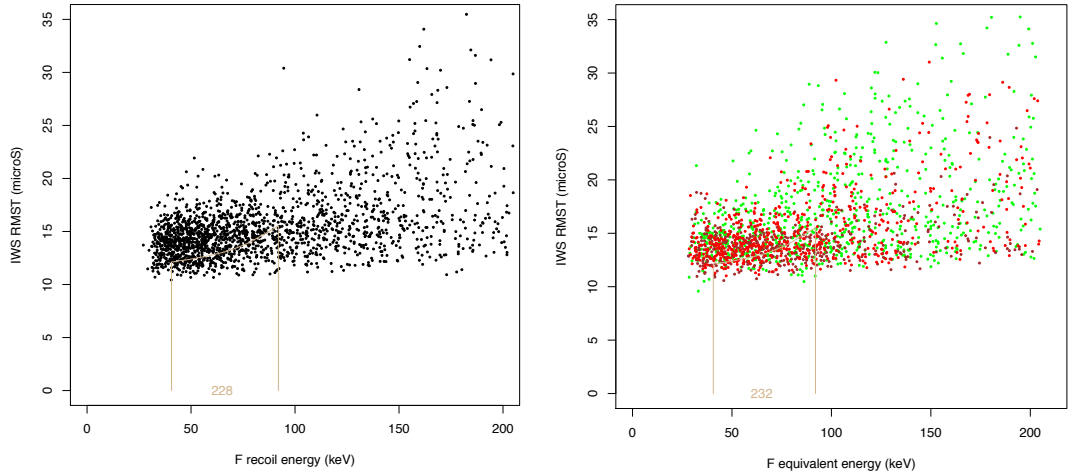

Fig. 4 – These plots show a comparison between real neutron data, on the left, and a *GEANT* + detector simulation of that same neutron exposure, on the right. For visualization purposes both data sets have the same number of accepted events, 1830. On the simulation plot the data are color coded red for F recoils, brown for S recoils and green for C recoils. Note that the number of events falling inside the acceptance window is identical within statistics.

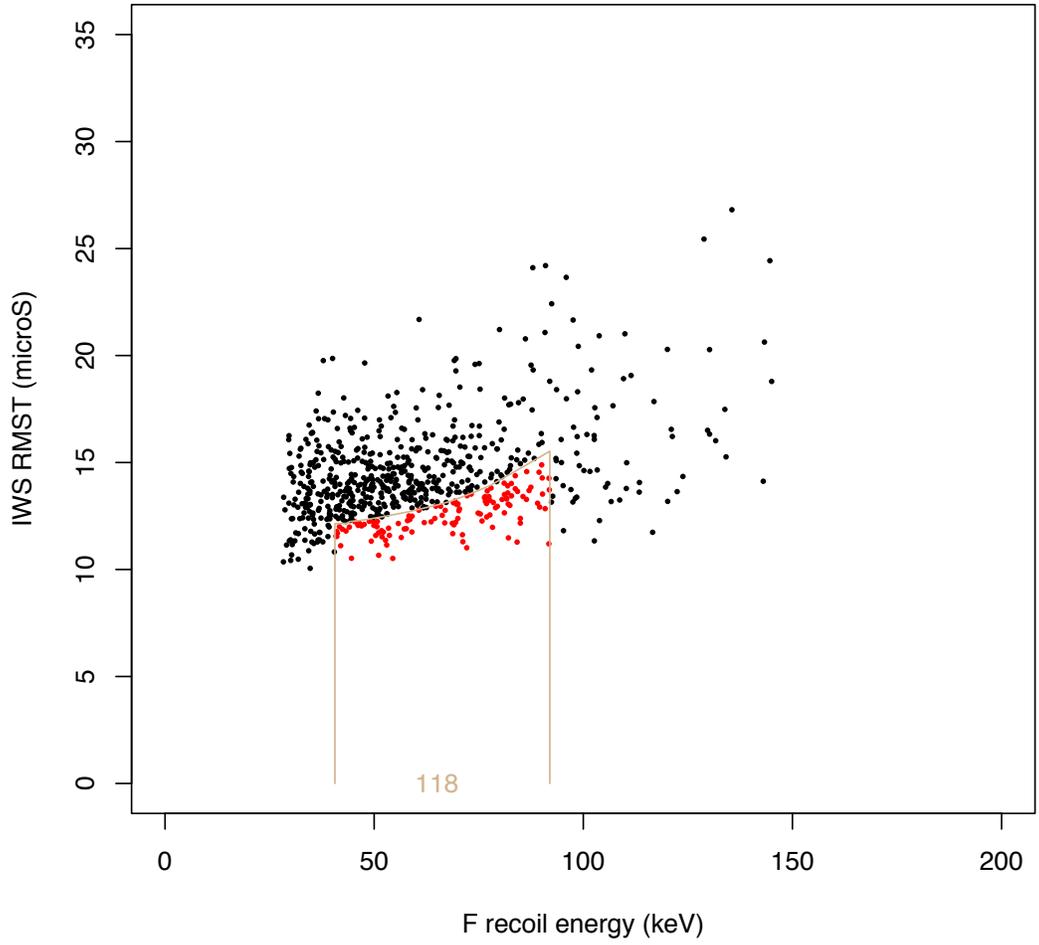

Fig. 5 – This plot shows the F recoil equivalent energy vs. *IWS RMST* for 100 GeV WIMPs. Events that fall within the acceptance window (tan line) are shown in red in the color plot. For reference the WIMP-proton cross section that would give this many events in 47.4 days of live time is 94 pb.

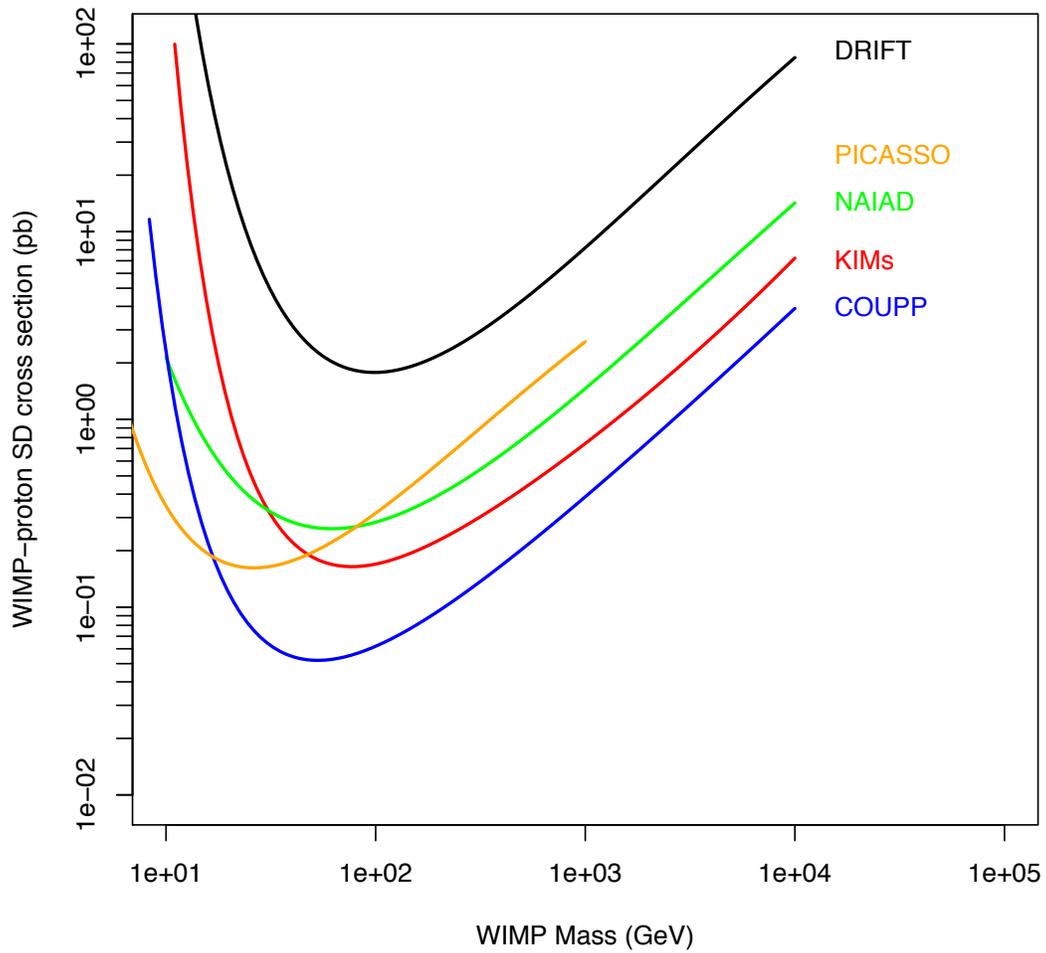

Fig. 6 – Spin-dependent WIMP-proton limits from the DRIFT-IId detector. DRIFT limits are shown in black while limits from PICASSO [18], NAIAD [20], KIMs [21], COUPP [22] and are shown in orange, green, red and blue. Note that DRIFT is the only directionally-sensitive experiment shown here.